# Architecture of a Silicon Strip Beam Position Monitor

**R. Angstadt[a], W. Cooper[a], M. Demarteau[a], J. Green[a], S. Jakubowski[a], A. Prosser[a], R. Rivera[a], M. Turqueti[a], M. Utes[a*], and Xiao Cai [b]**

[a] *Fermilab,*
*Wilson Rd. & Pine Street, Batavia, IL, USA*
[b] *Institute for High Energy Physics, Beijing, China*

*E-mail*: utes@fnal.gov

ABSTRACT: A collaboration between Fermilab and the Institute for High Energy Physics (IHEP), Beijing, has developed a beam position monitor for the IHEP test beam facility. This telescope is based on 5 stations of silicon strip detectors having a pitch of 60 microns. The total active area of each layer of the detector is about 12x10 $cm^2$. Readout of the strips is provided through the use of VA1` ASICs mounted on custom hybrid printed circuit boards and interfaced to Adapter Cards via copper-over-kapton flexible circuits. The Adapter Cards amplify and level-shift the signal for input to the Fermilab CAPTAN data acquisition nodes for data readout and channel configuration. These nodes deliver readout and temperature data from triggered events to an analysis computer over gigabit Ethernet links.

This work was supported by the U.S. Department of Energy, operated by Fermi Research Alliance, LLC under contract No. DE-AC02-07CH11359 with the United States Department of Energy.

## 1. Overview

The IHEP Silicon Test Beam position monitor consists of five layers of silicon detectors arranged to form a small beam telescope. Each layer (Figure 1) employs three 640-channel silicon strip detectors left over from the DZero Run IIb experiment at Fermilab, totaling 1920 silicon strips per layer. The strips are read out via wire-bond connections to VA1' ASICs, which are 128-channel "sample and hold" chips that simultaneously store all channels' charge when beam arrives, and hold each channel's charge for later readout. Readout occurs by applying a 2.5MHz bipolar clock, where each clock edge puts the next channel's differential current on the output pins. Per layer there are five VA1' chips residing on each of three Hybrid boards, which are mounted on a common frame with the detectors. The Hybrids also carry support components for the ASICs, including power supply bypassing, current sourcing for the preamplifiers and shapers, voltage biasing for the preamplifiers and shapers, and a temperature monitoring IC. Three hybrids connect, via 14cm long custom copper-over-polyimide cables, to the Adapter Card. The Adapter Card supplies +2V and -2V power to the ASICs, buffers the control signals from LVTTL to the +1.8V/-1.8V logic levels required by the ASICs, and translates the differential current output of the ASICs' analog output into the differential voltage needed by the downstream ADC. The downstream ADC resides on another board directly connected to the Adapter Card called CAPTAN (Compact And Programmable daTa Acquisition Node). CAPTAN is a versatile and expandable data acquisition and processing system which simply connects via Ethernet to a desktop computer. This board forms the control signals needed by the ASICs, digitizes and stores the analog data coming from the ASICs, and interfaces to the temperature readout IC on the Hybrid. All boards are designed by Fermilab for low noise and simplicity of connection.

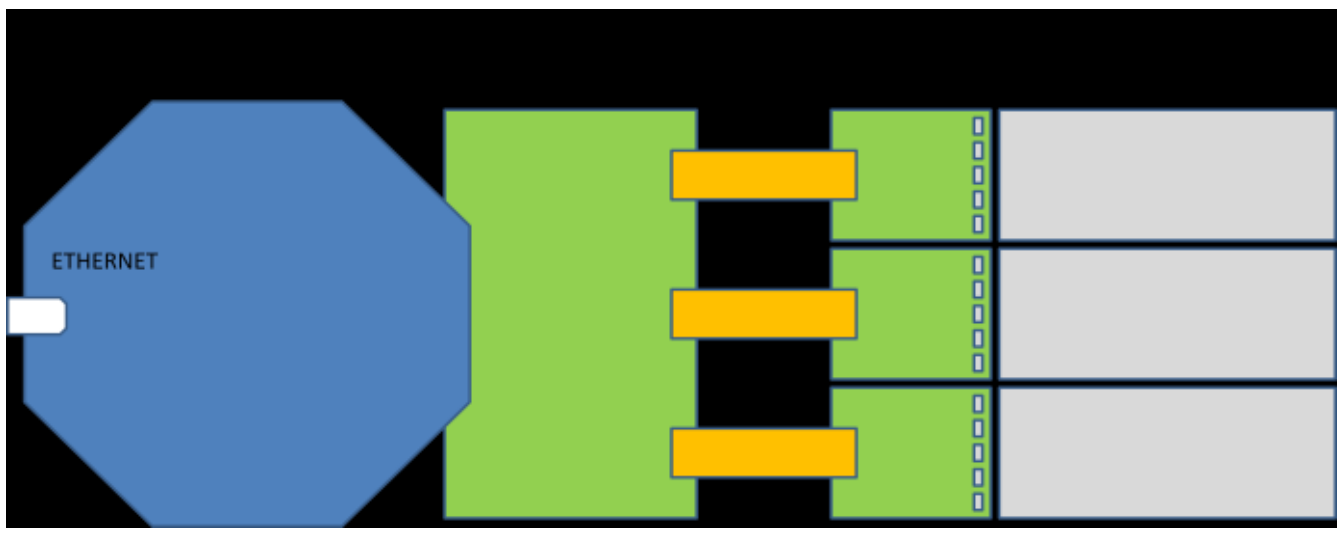

Figure 1. Block diagram of one layer of the IHEP test beam data acquisition system

## 2. Sensors and Mechanical Design

The sensors used for the IHEP project are silicon strip detectors remaining from Dzero's RunIIb project at Fermilab. The active length is 98.33 mm, the active width is 38.372 mm, and the thickness is 320μm. Strip pitch is 30μm, readout pitch is 60μm, and there are 639 channels per sensor. These sensors are AC-coupled, single-sided single-metal p+ on n- bulk silicon devices with integrated polysilicon resistors. The sensors are fully depleted with a nominal 120

Volts of bias. When a minimum-ionizing particle strikes the detector, the sensor produces electron-hole pairs and liberates about 3.5fC of charge per MIP. Leakage current is specified as <100nA/cm$^2$, junction breakdown >350V, 12pF/cm coupling capacitance, coupling capacitor breakdown >100V, and interstrip capacitance <1.2pF/cm.[1]

The hybrids are mounted directly onto the sensors (Fig. 2). A gold pad on the hybrid provides the bias voltage for the sensor. This structure, called a ladder, is mounted on an FR4 frame. The hybrids are mounted on top of the frame which has an embedded cooling channel to cool the ASICs. Cooling is provided by a 25% propylene glycol in distilled water mixture at 12$^o$C. The opposite side of the sensor is mounted directly on the passive side of the frame. Three ladders populate a single frame. The strips are oriented at 90$^o$ and +/- 5.71$^o$ for the first three planes and have interlayer spacing of 8 mm. The second set of two planes is located 6.6 cm downstream from the first set with strip orientation of 90$^o$ +/- 5.71$^o$.

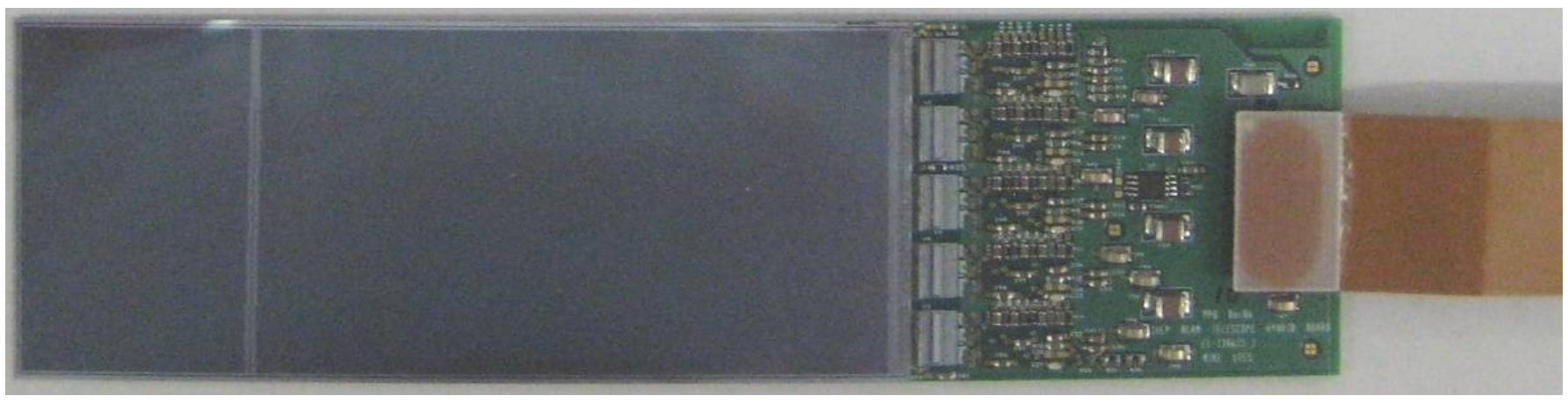

Figure 2. Sensor with hybrid and a partial view of the cable.

## 3. Hybrid

### 3.1 Mechanical Aspects

The hybrid (Fig. 3) is a six-layer FR4 circuit board that carries five VA1' ASICs and their associated passive components, a temperature sensing IC, and a connector for the flex cable that goes to the Adapter Card. Nominal dimensions are 40.3x55.5mm$^2$. There is no specification for controlled impedances due to the low frequencies involved; the main concern here is electric and magnetic field immunity and other noise reduction techniques.

### 3.2 VA1' ASICs and Associated Circuitry

The detector signals are fed via wire bonds into the VA1' ASIC[2] made by Ideas ASA in Norway. It is a 128 channel low-power charge sensitive preamp/shaper circuit. Sampling of the detector signal occurs on all channels at the falling edge of a Hold signal, which should occur after a "peaking time" of 0.5 to 1 microsecond following beam arrival time. Readout of the 128 channels occurs serially during application of a readout clock. Nominal power levels are +2V and -2V +/- 10% although we have found that the chips can work at voltages as low as +/- 1.4V. Input logic levels are approximately +/-1.8V. The analog output is a differential current with a specified gain of about 10μA/fC. Dynamic range is about 70fC in single polarity mode.

### 3.3 Temperature Monitoring

The temperature is monitored by Dallas Semiconductor DS28EA00 IC which measures from -40C to +85C, uses a 1-wire interface, and draws only 1.5μA standby current from the +3V bias supplied by the Adapter Card. Temperature information is included in the data stream. The 1-wire interface is sophisticated and is handled directly by firmware in the FPGA on the CAPTAN board.

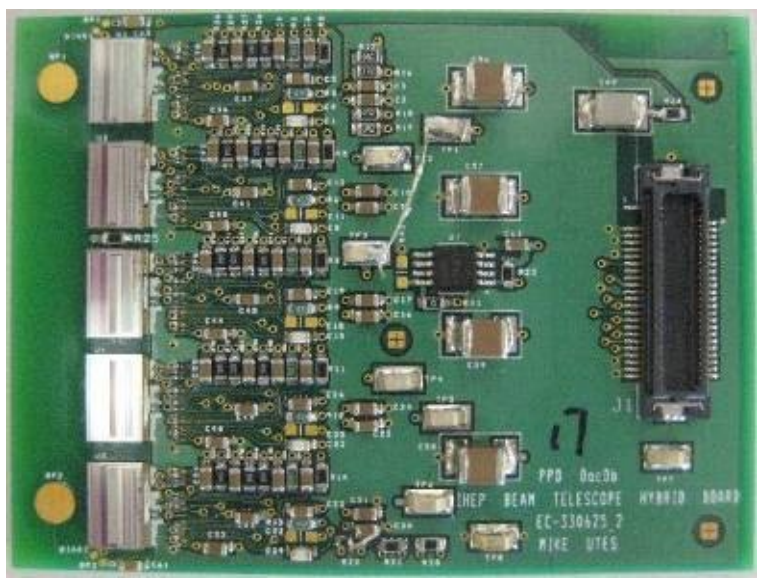

Figure 3. Hybrid showing VA1' ASICs and the gold pads at the far left for the bias voltage

## 4. Cable

The flex cable connects the Hybrid to the Adapter Card. Since the sensors and hybrids are in a temperature- and humidity-controlled enclosure, this cable allows for the rest of the readout electronics to be mounted outside the enclosure. The cable consists of two copper trace layers and kapton dielectric. One layer is a solid ground plane; all other signals and power are on the other layer. The cable has redundant vias to increase reliability. Center-to-center distance between connectors is 5.0" (12.7cm) and overall length is 5.5" (13.97cm); width is 0.8" (2.03cm). Since pin 1 carries the detector bias, pins 2, 3, 4 and 5 have been removed to eliminate the possibility of arcing.

## 5. Adapter Card

### 5.1 General

The IHEP Adapter Board (Fig. 4) interfaces three Hybrids to a CAPTAN board. This board measures 12.1 mm by 7.1 mm and is nominally 0.062" thick. There is one adapter board per layer.

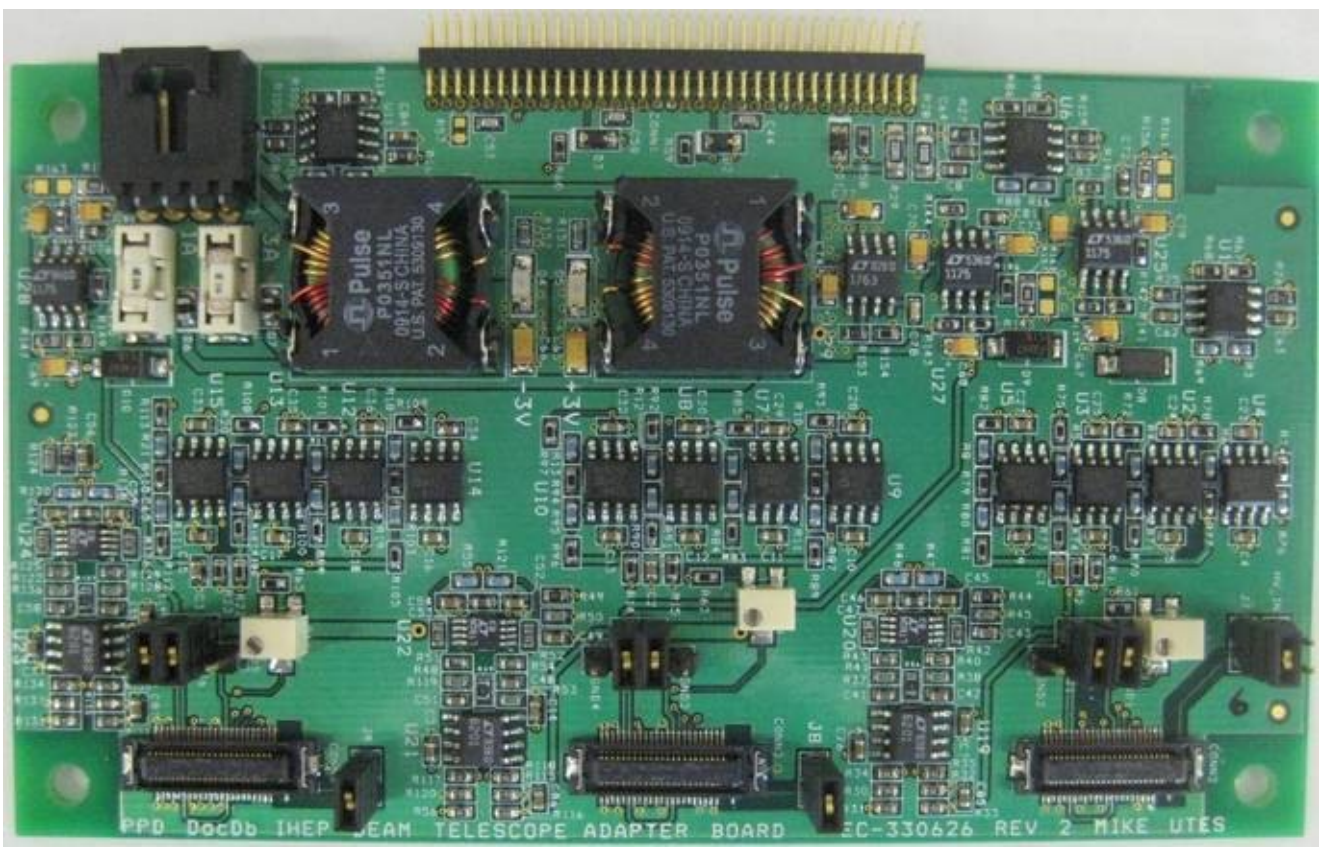

Figure 4. Adapter Card

### 5.2 Control Signal Buffering

The Adapter Board translates the control signals' LVTTL logic levels coming from CAPTAN into +1.8V high/ -1.8V low logic levels compatible with the VA1' chip. This is done with simple op-amp circuits that provide clean edges and low noise.

### 5.3 Differential Analog Readout Signal Circuitry

Another function is to translate the differential current levels coming from the VA1' analog output into differential voltage acceptable to the ADC on the CAPTAN board. The differential current from the VA1' ranges from -200mA to +200mA, and the adapter card translates this (Fig. 5) into a differential voltage within the limits of the ADC, which is specified as 0 to 1.4 Volts with a midpoint of 0.7V. The gain factor here is about 7.

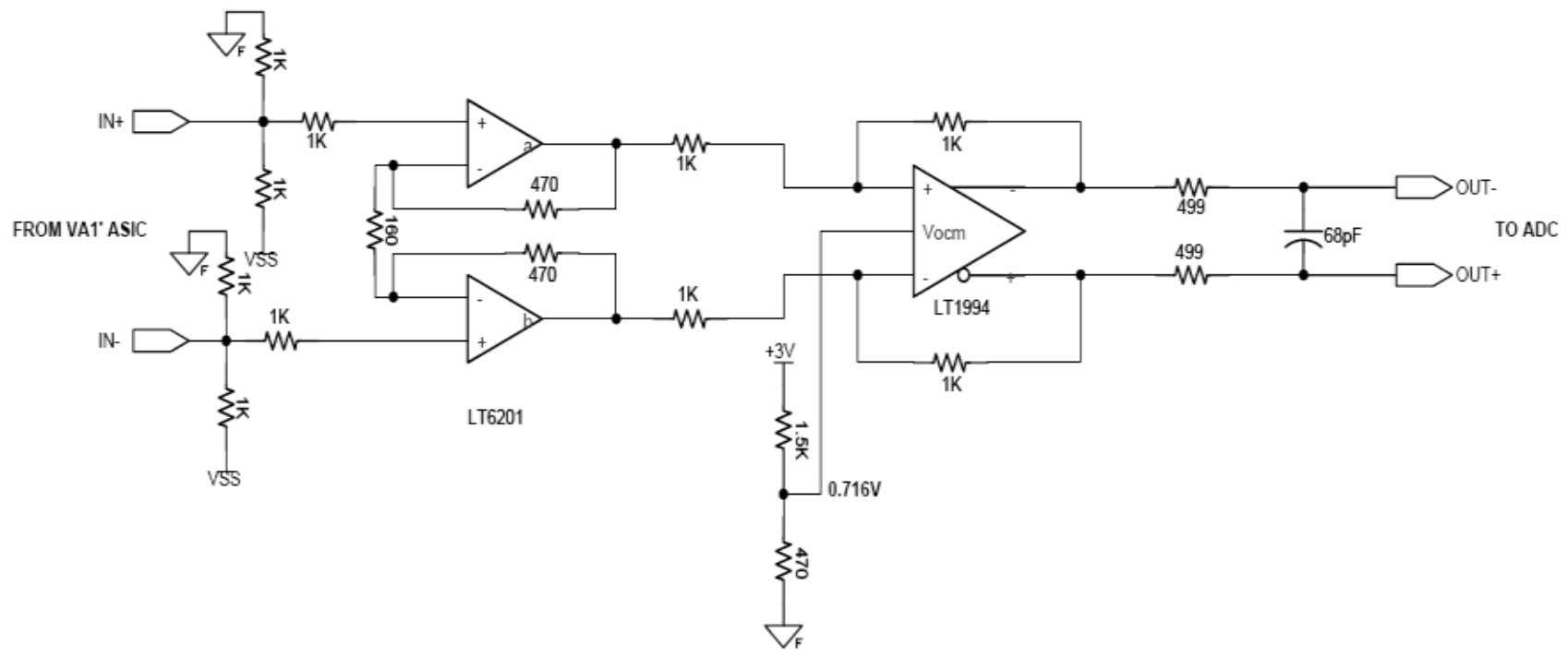

Figure 5. Adapter Card's Instrumentation amplifier with level shifting for matching to the ADC

### 5.4 Power Regulation for the Hybrid

Clean power is critical for the Viking ASICs on the Hybrid to keep the noise to an acceptable level, so we power them via linear regulators (LDOs). For providing +2V, there is one regulator shared by all three hybrids since the current demand is about 70mA. For the -2V supply there is one regulator for each Hybrid; the current draw is about 360mA.

## 6. CAPTAN

The CAPTAN system[3] is a flexible and expandable DAQ system interfaced to a PC via Ethernet. For this application, CAPTAN consists of two octagonal cards, a Data Conversion Board (DCB) and a Node Processing and Control Board (NPCB) stacked together (Fig. 6). There is one stack for each layer of the detector. From the PC, an Ethernet cable connects to the NPCB. A GUI on the PC defines the VA1' control waveforms which are stored in an FPGA on the NPCB. These control signals are passed down to the DCB, and flow through a 70-pin connector which mates to the Adapter Card. The VA1' acquires the data and the differential analog readout, buffered by the Adapter Card, is received by a 12-bit MAX1438 ADC on the DCB. The ADC sends digitized values back to the FPGA on the NPCB and formatted data then flows back to the PC via the Ethernet cable. The power supply requirement for CAPTAN is 3.5V, 1A. A small separate card allows for an external trigger, which is serially distributed to the other CAPTAN stacks though SATA cables.

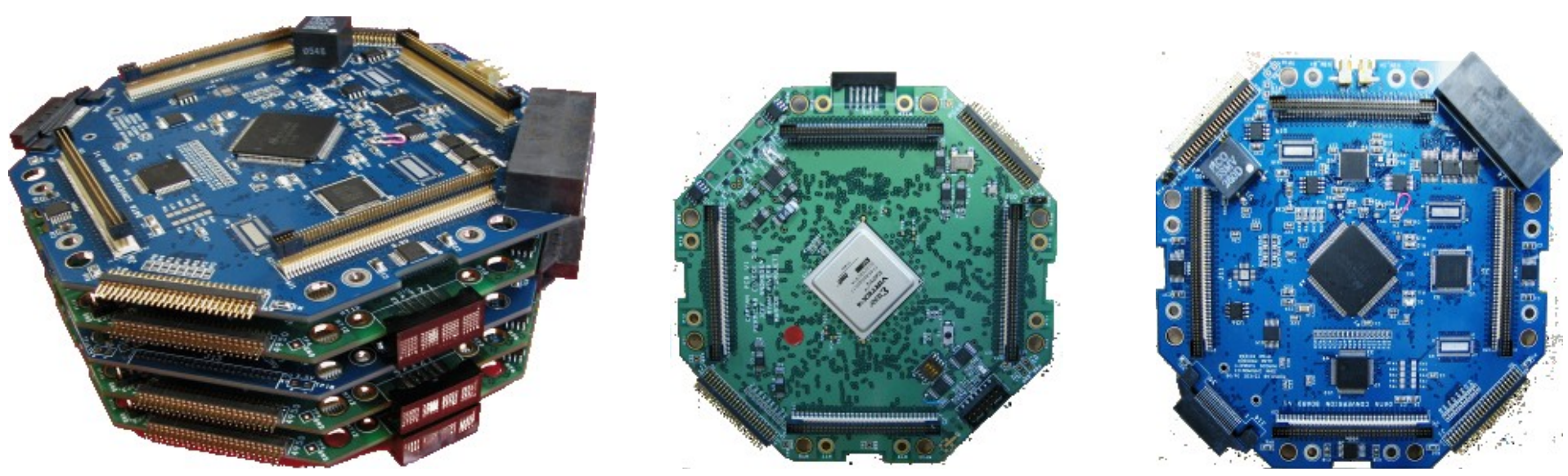

Figure 6. CAPTAN stack (left), Node Processing and Control Board (middle), and Data Conversion Board (right).

## 7. Results

The detector is now running in the IHEP test beam. The signal to noise ratio is greater than 20. This is expected since the equivalent noise charge (ENC) of the VA1' is 180 + 7.5*C (e) = 1080 (e) / strip. Expected S/N = 22 (for 300 μm Si). The differential noise is estimated[4] by computing the root mean square value of $(a_{n+1} - a_n)/\sqrt{2}$. In the data the differential noise is 20 ADC counts and the signal is typically greater than 400 ADC counts (Fig. 7).

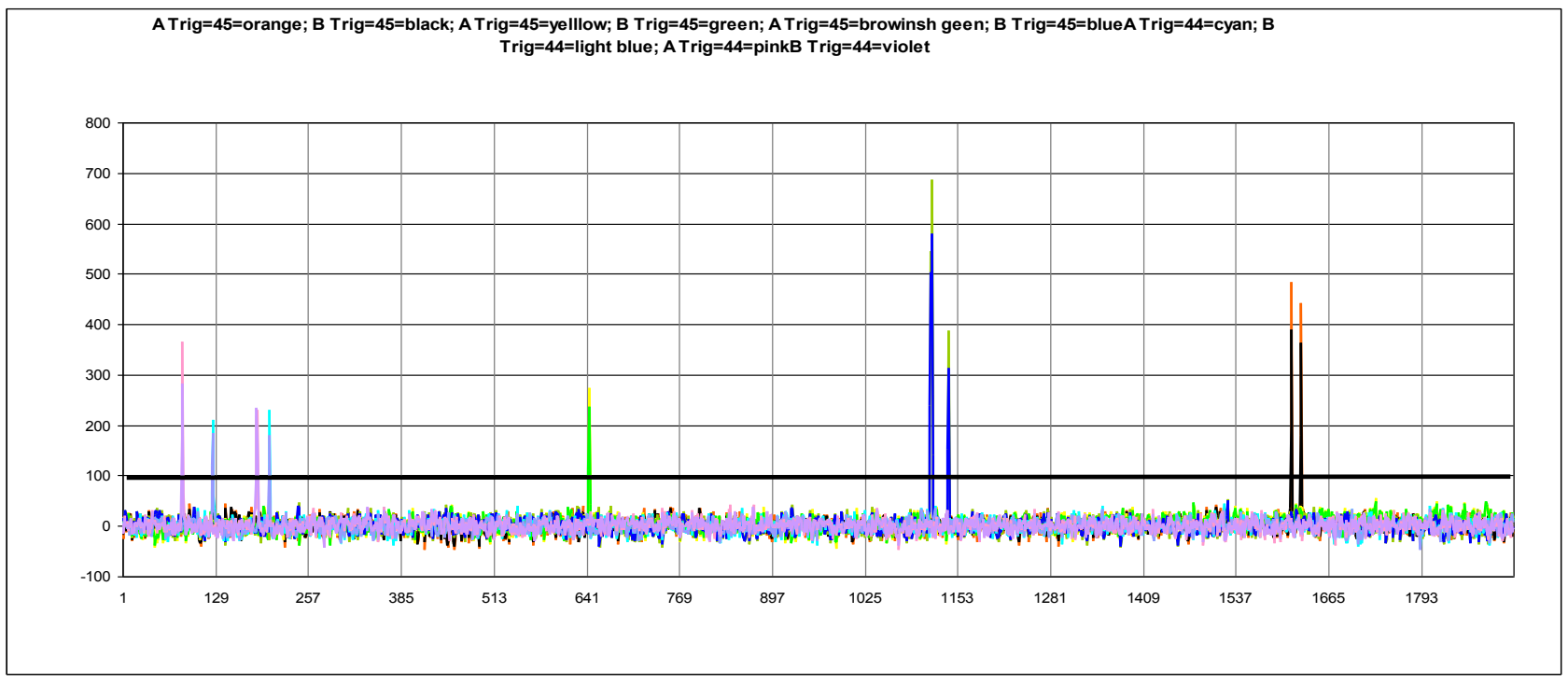

Figure 7. Pedestal subtracted data for an event that recorded the passing of two cosmic rays. Shown is the data for all 3x639 channels per layer superimposed on each other, differentiated by their color. Because of the different orientation of the various layers, the cosmic rays will register in different channels in different layers.

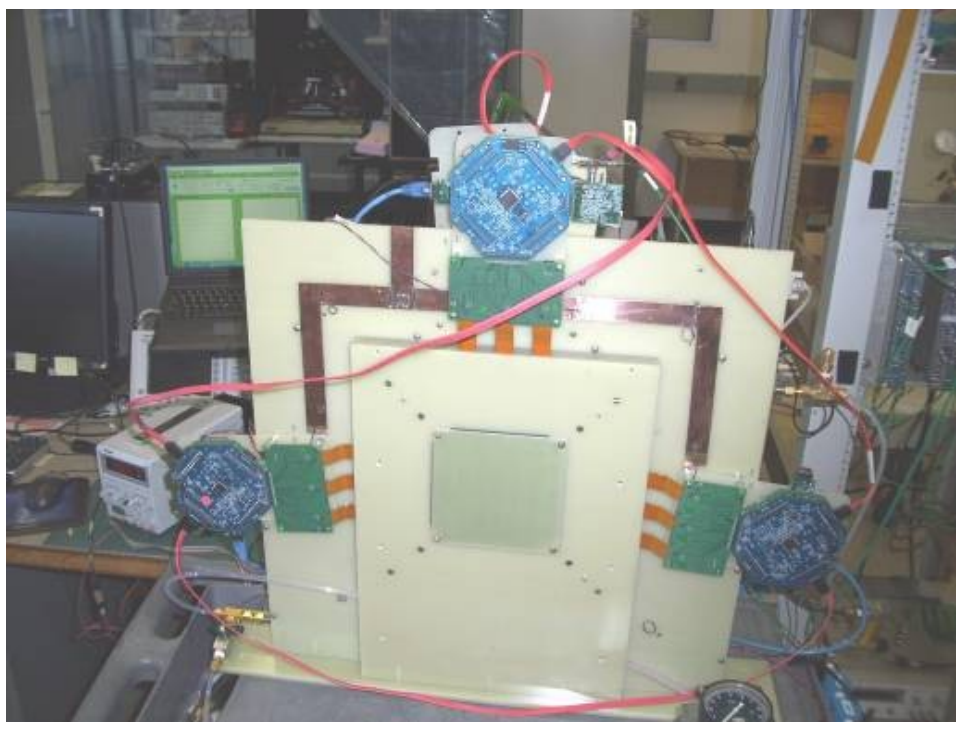

## Acknowledgments

We would like to thank Sergey Los of Fermilab for his help in improving our S/N ratio.